\documentclass[11pt]{article}
\usepackage{verbatim}
\usepackage{amsfonts}
\usepackage{graphics}
\usepackage{amsmath}
\usepackage{times}
\usepackage{appendix}
\usepackage{graphicx}
\usepackage{color}
\usepackage{enumerate}
\usepackage{fancyhdr,latexsym,amsmath,amsfonts,amssymb,amsbsy,amsthm,url}
\usepackage[hscale=0.7,centering]{geometry}
\usepackage{graphicx,epsfig}
\usepackage{amssymb,amsmath}
\usepackage{times}
\usepackage{color}
\usepackage{amsfonts}[11pt]
\usepackage{amsmath}[11pt]

\numberwithin{equation}{section}

\fancypagestyle{usual}{
  \lhead[\thepage]{}
  \chead[{\it \shortauthors}]{\sc \shorttitle}
  \rhead[]{\thepage}
  \lfoot[]{}
  \cfoot[]{}
  \rfoot[]{}
  
}

\makeatletter

\renewcommand{\section}{
  \@startsection
  {section}
  {1}
  {0pt}
  {1.1\baselineskip}
  {0.2\baselineskip}
  {\sc \centering}
}

\renewcommand{\subsection}{
  \@startsection
  {subsection}
  {1}
  {0pt}
  {1.1\baselineskip}
  {0.2\baselineskip}
  {\sc \centering}
}

\renewcommand{\subsubsection}{
  \@startsection
  {subsubsection}
  {1}
  {0pt}
  {1.1\baselineskip}
  {0.2\baselineskip}
  {\sc \centering}
}

\makeatother

\begin{document}

\title{\large\sc Optimal Therapy of Hepatitis C Dynamics and Sampling Based Analysis
}\normalsize
\author{\sc{Gaurav Pachpute} \thanks{Department of Mathematics, Indian Institute of Technology Guwahati, Guwahati 781039, Assam, India,
e-mail: g.pachpute@iitg.ernet.in}
\and \sc{Siddhartha P. Chakrabarty} \thanks{Department of Mathematics, Indian Institute of Technology Guwahati, Guwahati 781039, Assam, India,
e-mail: pratim@iitg.ernet.in}}
\date{}
\maketitle
\begin{abstract}
We examine two models for hepatitis C viral (HCV) dynamics, one for monotherapy with interferon (IFN) and the other for
combination therapy with IFN and ribavirin. Optimal therapy for both the models is determined using the steepest gradient
method, by defining an objective functional which minimizes the infected hepatocyte levels, virion population and the
side-effects of the drug(s). The optimal therapy for both the models shows an initial period of high efficacy,
followed by a gradual decline. The period of high efficacy coincides with a significant decrease in the
infected hepatocyte levels as well as viral load, whereas the efficacy drops after liver regeneration through restored
hepatocyte levels. The period of high efficacy is not altered significantly when the cost coefficients are varied, as long as
the side effects are relatively small. This suggests a higher dependence of the optimal therapy on the model
parameters in case of drugs with minimal side effects.

We use the Latin hypercube sampling technique to randomly generate a large number of patient scenarios (i.e, model
parameter sets) and study the dynamics of each set under the optimal therapy already determined. Results show an increase in
the percentage of responders (as indicated by drop in viral load below detection levels) in case of combination therapy as
compared to monotherapy. Statistical tests performed to study the correlations between sample parameters and the time
required for the viral load to fall below detection level, show a strong monotonic correlation with the death rate of
infected hepatocytes, identifying it to be an important factor in deciding individual drug regimens.

\textbf{Keywords:} Hepatitis C, Optimal Control, Steepest-Gradient Method, Latin Hypercube Sampling, Statistical Tests

\end{abstract}

\def\Jc{\mathcal{J}}
\def\Hc{\mathcal{H}}
\def\Lc{\mathcal{L}}
\def\Kc{\mathcal{K}}
\def\xbf{\mathbf{x}}
\def\ubf{\mathbf{u}}
\def\fbf{\mathbf{f}}
\def\lambf{\mathbf{\lambda}}
\def\Jbf{\mathbf{J}}
\def\SS{\scriptsize}
\def\TM{\!\!$^{\mbox{\texttt{\SS TM}}}$~}

\section{Introduction}

Hepatitis C (HCV) is an infectious disease which spreads through blood contact. It is estimated that HCV has infected 170
million individuals worldwide \cite{Dixit08}. Roe and Hall \cite{Roe08} estimate that about $50-80 \%$ of HCV infected
cases are chronic in nature. Of these chronic cases, about $10-20 \%$ develop into liver cirrhosis of which, about $5 \%$
develop hepatocellular carcinoma (HCC). The extend of prevalence of HCV varies widely across geographical locations
\cite{Razali07,Castello10}. While the dominant mode of HCV transmission in United States, Europe and Australia is injecting drug use
\cite{Razali07}, the absence of reliable screening for HCV amongst blood donors remains a major challenge in combating the spread of
the disease in India \cite{Pal02,Mukhopadhya08}.

Mathematical modelling and quantitative analysis of hepatitis C infections has been explored
extensively over the last decade. Most of the modelling has been restricted to the short term dynamics of the model.
One of the earliest models was proposed by Neumann et al. \cite{Neumann98}, who examine the dynamics of HCV
in presence of Interferon-$\alpha$ (IFN-$\alpha)$ treatment. They find that the primary role of IFN is in blocking
the production of virions from the infected hepatocytes. However, IFN has little impact when it comes to controlling the
infection of the hepatocytes. Dixit et al. \cite{Dixit04} improved upon \cite{Neumann98} by including the effects of ribavirin ,
which in turn results in a fraction of the virions being rendered noninfectious. Their model is able to explain clinically observed
biphasic decline patterns amongst patient population. Their study also shows that while IFN plays a pivotal role in the first phase
decline of viral load, ribavirin has very little impact.
However, in case of low IFN efficacy, ribavirin makes a significant contribution to the second phase of decline.
The model could not successfully explain the triphasic decline patterns, as well as some cases of non-responders.
Dahari et al. \cite{Dahari07} in a subsequent and improved model, take into account the homeostatic mechanisms for the liver by incorporating
a growth function. This model successfully explains the triphasic decline, as well as therapeutic failures.

Control theory has found wide ranging applications in biological and ecological problems \cite{LenhartBook}.
In biomedical problems, techniques from control theory are of great use in developing optimal therapeutic strategies.
The treatment regimen is usually taken to be the control variable, with the aim of minimizing the detrimental effects of the
medical condition. For instance, in cancer modeling, DePillis et al. \cite{DePillis07} examine a mathematical model of tumor–-immune interactions
and determine the optimal chemotherapy with the goal of minimizing the tumor density, as well as the therapeutic side effects,
for both linear and quadratic control. Fister and Panetta \cite{Fister03}
consider three different cell-kill models and determine the optimal drug dosage which minimizes the cancer mass and the cost
of treatment for all three models. Murray \cite{Murray90} uses linear control approach for cancer chemotherapy models with toxicity limit.
Optimal control theory can be applied to epidemic models, such as the SEIR model, with the goal of minimizing the number of
infectious individuals and the overall cost of the vaccination programme \cite{LenhartBook}. In case of treatment of diabetes,
the control is taken as the insulin injection levels with the objective of minimizing the difference between current and the
desired glucose levels. Chavez et al. \cite{Chavez09} present an optimal insulin delivery for type 1 diabetic patients.

Control theoretic approach has been applied extensively in case of virological models, especially in case of HIV models.
One of the earliest papers in this area by Kirschner et al. \cite{Kirschner97} uses an existing ODE model which incorporates the
dynamics of the immune system and HIV. Using an objective functional which maximizes the levels of T-cells and minimizes the cost
of treatment, the optimal chemotherapy (like protease inhibitors) is determined for various stages of initiation of treatment.
Joshi \cite{Joshi02} considers a model with combination therapy and numerically solves the optimal system for the optimal
treatment regimen. Adams et al. \cite{Adams05} present the optimal treatment protocols (along with modeling and data analysis)
for HIV dynamics. Stengel \cite{Stengel08}, in his paper, obtains the optimal therapies for drug resistant strains of HIV which evolve
rapidly, as a result of high viral turnover and mutation rates. He observes that in such cases continued treatment is imperative for
sustained remission.

In the case of HCV, Chakrabarty and Joshi \cite{SPCJBS} consider a model
(motivated by \cite{Neumann98,Dixit04,Dahari07})
for HCV dynamics under combination therapy of interferon and ribavirin. An objective functional is formulated
to minimize the viral load, as well as the drug side-effects and the optimal system is solved numerically to
determine optimal efficacies of the drugs. Chakrabarty \cite{SPCOCAM} extended the results in \cite{SPCJBS}
by considering a clinically validated functional form for the interferon efficacy and hence determined the optimal
efficacy of ribavirin. Martin et al. \cite{Martin11} in a recent paper examine a three compartment model for HCV,
involving the susceptible, chronically infected and treated injecting drug users (IDUs). They determine an optimal
treatment programme over a 10 year period taking into account several biomedical and economic objectives.

\section{Model for Monotreatment with IFN}

We first consider a three dimensional model given by Dahari et al. \cite{Dahari07} based on an earlier
pioneering model of Neumann et al. \cite{Neumann98}. The model describes the dynamics of the
the uninfected and infected hepatocytes as well as the HCV and is expressed as a system of three coupled
ordinary-differential equations:
\begin{eqnarray}
\label{model1eqn}
\frac{dT}{dt}&=&s+rT\left(1-\frac{T+I}{T_{\max}}\right)-dT-\beta T V \nonumber\\
\frac{dI}{dt}&=&\beta TV+rI\left(1-\frac{T+I}{T_{\max}}\right)-\delta I \nonumber\\
\frac{dV}{dt}&=&(1-\epsilon_{p})p I-cV.
\end{eqnarray}
The uninfected hepatocytes ($T$) are being produced at a rate $s$ and proliferate logistically at a rate $r$,
accompanied by a natural death rate of $d$. The homeostatic liver mechanism is incorporated through $T_{\max}$,
which is the maximum hepatocyte count in the liver.
HCV ($V$) is assumed to infect the hepatocytes at a rate $\beta$, thereby producing infected hepatocytes ($I$).
The infected hepatocytes are also assumed to proliferate logistically at a rate $r$ and have a natural death rate of $\delta$.
In the absence of any kind of treatment, the infected hepatocytes produce HCV at a rate $p$, which has a clearance rate of
$c$. The production of HCV is lowered by a factor of $(1-\epsilon_{p})$ upon the administration of IFN treatment,
where $\epsilon_{p}$ is the efficacy of IFN. The model excludes the role of IFN in blocking the infection, as
it was observed to have minimal effect in this case \cite{Dixit08,Dixit04}.
This model admits two steady states \cite{Dahari07}, namely the uninfected state,
\[T^{(u)}= \frac{T_{\max}}{2r}\left[r-d+\sqrt{(r-d)^{2}+\frac{4rs}{T_{\max}}}\right],
I^{(u)}=0, V^{(u)}=0,\]
and the infected state,
\[T^{(i)}=\frac{1}{2}\left[-D+\sqrt{D^{2}+\frac{4sT_{\max}}{rA^{2}}}\right],
I^{(i)}=T^{(i)}(A-1)+T_{\max}-B,
V^{(i)}=\frac{(1-\epsilon_{p})pI^{(i)}}{c},\]
where,
\[A=\frac{(1-\epsilon_{p})p\beta T_{\max}}{cr}, B=\frac{\delta T_{\max}}{r},
D=\frac{1}{A}\left[T_{\max}+\frac{dB}{\delta A}-B\left(\frac{1}{A} + 1\right)\right].\]
Under the physiological conditions $r>d$ and $s\le dT_{\max}$, it can be shown \cite{Dahari07}
that there is a transcritical bifurcation at,
\[(1-\epsilon_{p})=\frac{c(\delta T_{\max} + r T^{(u)}-r T_{\max})}{p\beta  T^{(u)} T_{\max}}=C_{1}^{*} (\text{say}).\]
In other words, the uninfected steady state is stable if $(1-\epsilon_{p})<C_{1}^{*}$, while the
infected steady state is stable if $(1-\epsilon_{p})>C_{1}^{*}$.

\section{Optimal Control Problem for Monotreatment Model}

In this section, we present an optimal control problem motivated by biomedical considerations.
The goal is to formulate the problem such that, the integral and terminal values of HCV ($V$), as well as that of
the infected hepatocytes ($I$) are minimized with respect to the control ($\epsilon_{p}$) (on the lines of \cite{Stengel08}).
The problem also incorporates the necessity of minimizing the therapeutic side effects of IFN.
Keeping these goals in mind, the objective functional, to be minimized, is defined as,
\[\Jc=\frac{1}{2}\left[ S_{22}I(t_{f})^2+S_{33}V(t_{f})^2 \right]
+\frac{1}{2}\int_{t_{i}}^{t_{f}}\left[Q_{22}I(t)^2+Q_{33}V(t)^2 + R\epsilon_{P}(t)^{2}\right]dt, \]
where, $S_{ii}$ and $Q_{ii}$ represent the cost coefficient associated with the respective variables.
For the purpose of notational convenience, we define the state variables as $x_{1}=T, x_{2}=I, x_{3}=V$
and the control variable as $u=\epsilon_{p}$.
The state equations in the new variables are given by,
\begin{eqnarray}
\label{model1eqnnew}
\dot{x}_{1}&=&s+rx_{1}\left(1-\frac{x_{1}+x_{2}}{T_{\max}}\right)-dx_{1}-\beta x_{1}x_{3}\nonumber\\
\dot{x}_{2}&=&\beta x_{1} x_{3}+rx_{2}\left(1-\frac{x_{1}+x_{2}}{T_{\max}}\right)-\delta x_{2}\nonumber\\
\dot{x}_{3}&=&(1-u)p x_{2}-cx_{3}.
\end{eqnarray}
By defining the state vector $\xbf=\begin{pmatrix}x_{1}&x_{2}&x_{3}\end{pmatrix}^{\top}$, the above equations
can be written as $\dot{\xbf}=\fbf(\xbf(t),u(t),t)$.
The initial condition, $\xbf(t_{i})=\xbf_{0}$, is taken to be the infected steady state concentrations
before the initiation of the treatment (when $\epsilon_{p}=u=0$)  \cite{Dahari07}.

The above problem can be viewed as a part of a more general setting as described below \cite{Stengel08,Kirk04}:\\
To find an admissible optimal control $\ubf^{*}:[t_{i},t_{f}]\rightarrow[\underline{\ubf},\overline{\ubf}]$,
which for the system $\dot{\xbf}(t)=\fbf(\xbf(t),\ubf(t),t), \xbf(t_{i})=\xbf_{0}$,
minimizes the cost functional
\begin{eqnarray*}
\Jc&=&\frac{1}{2}\xbf^{\top}(t_{f})S\xbf(t_{f})+ \int_{t_{i}}^{t_{f}} \frac{1}{2}\left[\xbf^{\top}(t)Q\xbf(t) + \ubf^{\top}(t)R \ubf(t)\right]\\
&=&\Kc\left(\xbf(t_{f}),t_{f}\right)+\int_{t_{i}}^{t_{f}} \Lc(\xbf(t),\ubf(t),t)dt.
\end{eqnarray*}
The \textit{Hamiltonian}, defined by using the dynamic constraint $\fbf(\xbf(t),\ubf(t),t)$ and the Lagrangian
$\Lc(\xbf(t),\ubf(t),t)$ through the adjoint vector $\lambda^{T}$ (with the same dimension as the state vector), is as follows,
\[\Hc(\xbf(t),\ubf(t),\lambda(t),t)=\Lc(\xbf(t),\ubf(t),t)+\lambda^{T}(t)\fbf(\xbf(t),\ubf(t),t).\]
Using Pontryagin's minimum principle, the necessary conditions (in terms of the Hamiltonian) for $\ubf^{*}$ to be an
optimal control are \cite{Stengel08,Kirk04},
\begin{enumerate}[(a)]
\item $\dot{\xbf}^{*}(t)=\nabla_{\lambda}\Hc(\xbf^{*}(t),\ubf^{*}(t),\lambda^{*}(t),t)=\fbf(\xbf^{*}(t),u^{*}(t),t)$.
\item $\xbf^{*}(t_{i})=\xbf_{0}$.
\item $\dot{\lambda}^{*}(t)=-\left[\nabla_{\xbf}\Hc(\xbf^{*}(t),\ubf^{*}(t),\lambda^{*}(t),t)\right]^{\top}=
-\left[\nabla_{\xbf}\Lc(\xbf^{*}(t),\ubf^{*}(t),t)\right]^{\top}
-\left[\nabla_{\xbf}\fbf(\xbf^{*}(t),\ubf^{*}(t),t)\right]^{\top}\lambda(t)$.
\item $\lambda^{*}(t_{f})=\nabla_{\xbf}\Kc\left(\xbf^{*}(t_{f}),t_{f}\right)=S \xbf^{*}(t_{f})$.
\item $\nabla_{\ubf}\Hc(\xbf^{*}(t),\ubf^{*}(t),\lambda^{*}(t),t)=
\nabla_{\ubf}\Lc(\xbf^{*}(t),\ubf^{*}(t),t)
+\lambda(t)^{\top}\nabla_{\ubf}\fbf(\xbf^{*}(t),\ubf^{*}(t),t)=0$.
\end{enumerate}
The optimal control system thus, comprises of a coupled forward state equation and a backward adjoint equation, along with the
regular control. This problem, being nonlinear and coupled in nature, needs to be solved using concurrent and iterative
numerical procedures. In this paper, the solution is approximated by the steepest gradient method \cite{Stengel08,Kirk04}.
\begin{enumerate}
\item Approximation starts with an initial guess for the control $\ubf(t)=\ubf^{(0)}(t)$.
\item Using this control $\ubf^{(0)}(t)$, $\xbf^{(0)}(t)$ is found by numerically solving $\dot{\xbf}=\fbf(\xbf(t),\ubf^{(0)}(t),t),
\xbf(t_{i})=\xbf_{0}$, which is then used to solve
$\dot{\lambda}=-\nabla_{\xbf}\Hc(\xbf^{(0)}(t),\ubf^{(0)}(t),\lambda(t),t),
\lambda(t_{f})=S \xbf^{(0)}(t_{f})$ to obtain $\lambda^{(0)}(t)$.
\item The gradient $\nabla_{\ubf}\Hc(\xbf^{(0)}(t),\ubf^{(0)}(t),\lambda^{(0)}(t),t)$ is evaluated to obtain the updated control as follows,
\[\ubf^{(1)}(t)=\ubf^{(0)}(t)-\tau(0)\nabla_{\ubf}\Hc(\xbf^{(0)}(t),\ubf^{(0)}(t),\lambda^{(0)}(t),t)\]
\end{enumerate}
The above iterative process for the $k+1$-th iteration is given by,
\[\ubf^{(k+1)}(t)=\ubf^{(k)}(t)-\tau(k)\nabla_{\ubf}\Hc(\xbf^{(k)}(t),\ubf^{(k)}(t),\lambda^{(k)}(t),t)\]
This procedure is repeated until some convergence criterion, such as
$||\ubf^{(k+1)}(t)-\ubf^{(k)}(t)||<\text{tol}$ or $\Jc^{(k)}-\Jc^{(k+1)}<\text{tol}$, is satisfied (the latter was used in this paper).
Runge-Kutta method of order 4 is used for numerically integrating $\xbf$ and $\lambda$.

{\SS
\begin{center}
\begin{table}[h!]
\begin{center}
\begin{tabular}{|c|c|}
\hline
Parameter& Value (\cite{Dahari07}\\
\hline
$s$ & $1.0$ cell ml$^{-1}$ day$^{-1}$\\
$d$ & $0.01$ day$^{-1}$\\
$p$ & $2.9$ virions day$^{-1}$\\
$\beta$ & $2.25\times 10^{-7}$ ml day$^{-1}$ virions$^{-1}$\\
$c$ & $6.0$ day$^{-1}$\\
$\delta$ & $1.0$ day$^{-1}$\\
$r$ & $2.0$ day$^{-1}$\\
$T_{\max}$ & $3.6\times 10^{7}$ cells ml$^{-1}$\\
\hline
\end{tabular}
\caption{Parameter values for the optimal control simulation
\label{tableone}}
\end{center}
\end{table}
\end{center}
}

\section{Results and Discussion for Monotreatment Model}

We implemented the above procedure using MATLAB $^{\mbox{\texttt{\SS TM}}}$.
The optimal control is approximated using the steepest gradient method for the parameter values
given in Table \ref{tableone} \cite{Dahari07}. The efficacy $\epsilon_{p}$ can theoretically lie between $0$ and $1$. Accordingly, $\underline{u}$ is set to be
$0$, which corresponds to zero effectiveness of the drug. The value of $\overline{u}$, however, is taken to be less than $1$, for
biomedical observations suggest that a perfect efficacy is unlikely to be achieved.
A more mathematical justification can be presented using pharmacokinetic models (such as
Powers et al. \cite{Powers03}), where the term for efficacy cannot become $1$ for all possible values of drug doses.

\begin{figure}[hb]
\centering
\includegraphics[width=0.7\textwidth]{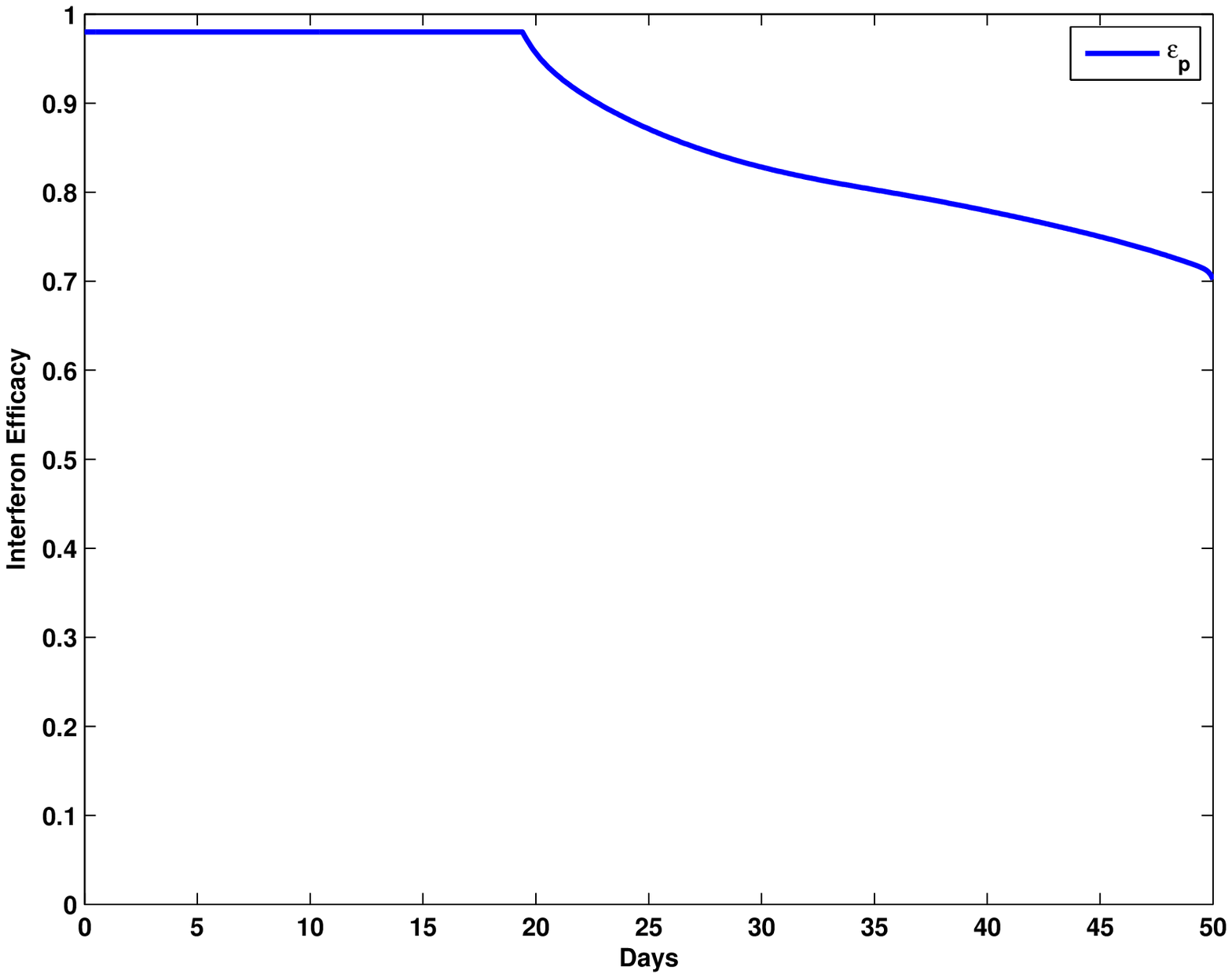}
\caption{Optimal Control for Monotherapy}
\label{three_control}
\end{figure}

The optimal control ($u^{*}$) is simulated over a period of 50 days and is presented in Figure \ref{three_control}. The cost
coefficients used were $S_{22}=S_{33}=10^{-5},Q_{22}=Q_{33}=10^{-5}$ and $R=0.1$.
It can be observed from the figure that, the optimal IFN efficacy remains at $\overline{u}=0.98$ for about three weeks and gradually decreases to about
$0.7$ over the next $4$ weeks. The same simulation, run over a period of ten weeks, exhibits a slower rate of decline in the optimal
efficacy. However, the time window for which the optimal efficacy remains at $\overline{u}$, is roughly unchanged.
The simulations are repeated for various values of matrices $S$, $Q$ and $R$, for various treatment time windows.
The results indicate that in responsive patients, the length of this period of administering high dosage does not vary by much
as long as the cost coefficient $R$ is relatively small. Higher values of $R$ result (as one expects) in a shorter period
of high dosage. This is of crucial importance when deciding on an optimal therapy of drugs with minor side-effects.

\begin{figure}[hb]
\centering
\includegraphics[width=0.7\textwidth]{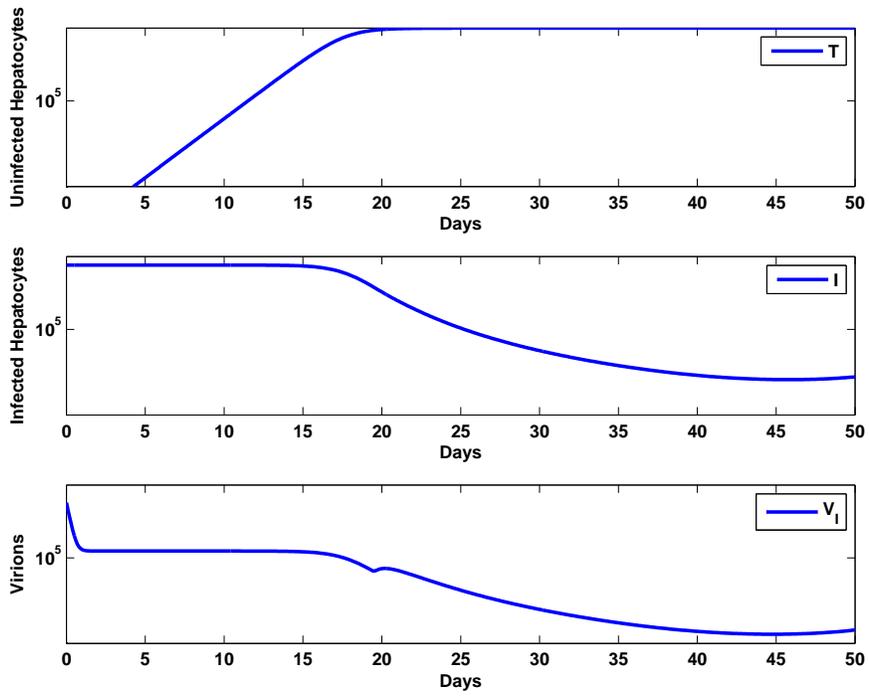}
\caption{State Dynamics for Monotherapy}
\label{three_state}
\end{figure}

\begin{figure}[hb]
\centering
\includegraphics[width=0.7\textwidth]{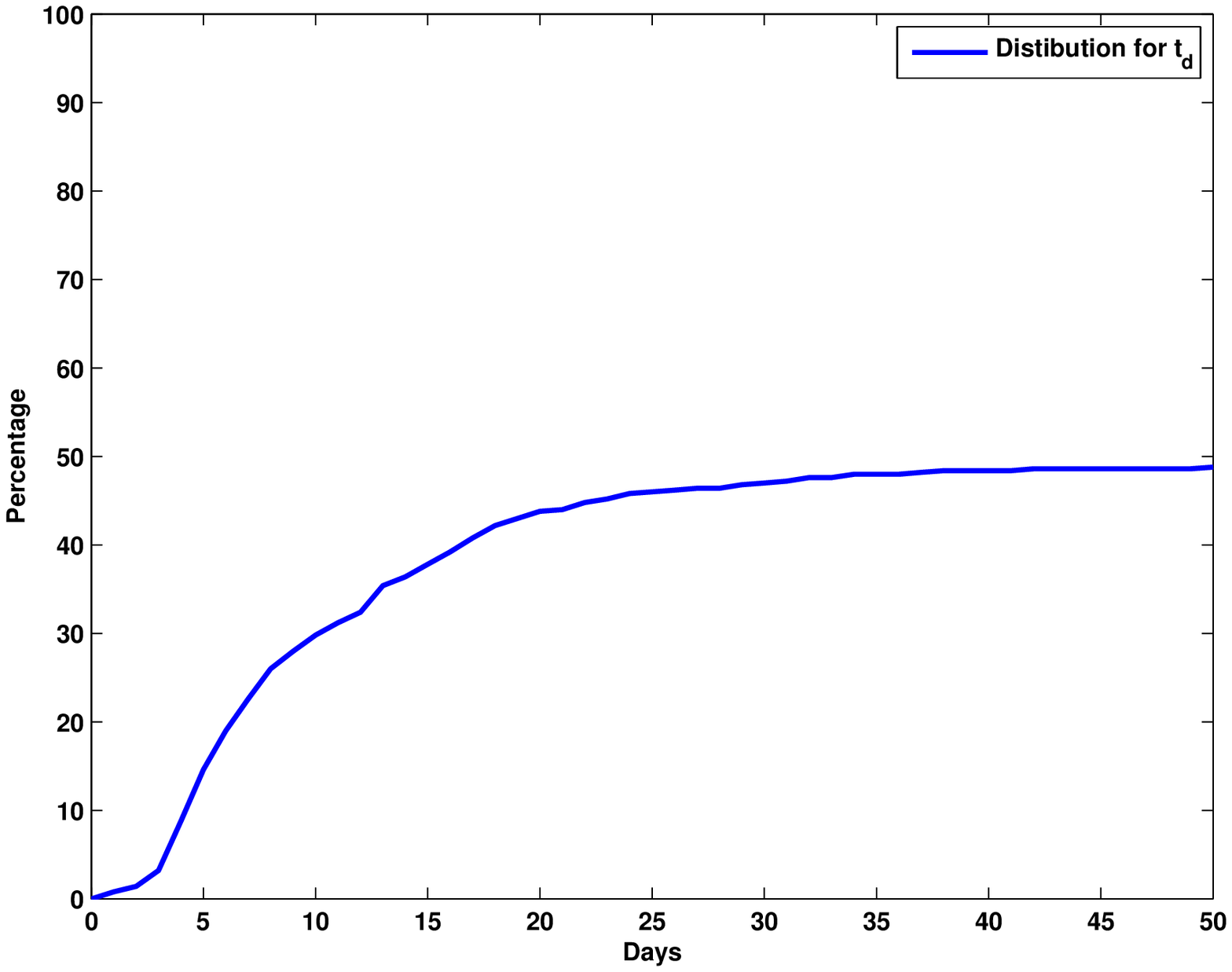}
\caption{Distribution of $t_{d}$ for Monotherapy}
\label{three_distribution}
\end{figure}

The dynamics of the uninfected ($T$) and infected ($I$) hepatocyte concentrations and HCV ($V$) levels
under the application of optimal efficacy (Figure \ref{three_control}) are given in Figure \ref{three_state}.
The viral load shows a triphasic decline, which is in line with the biomedical observations \cite{Dahari07}.
The optimal treatment exhibits a rapid first phase decline in the viral load during the first couple of days, followed by
a stable concentration over the next couple of weeks and finally a faster decline for the rest of the treatment period.
These observations are consistent with the results in \cite{Neumann98,Dixit04,Dahari07}.
The infected hepatocyte concentration decreases rapidly from the third week of the treatment, whereas, the uninfected hepatocyte concentration
increases over the course of the first three weeks of therapy and remains close to $T_{\max}$ (maximum liver volume) for the rest of the period.
Interestingly, the level of optimal efficacy remains at $\overline{u}$ until a point (three weeks in this case) where the viral load and the
infected hepatocyte count starts showing a rapid decline, accompanied by the uninfected hepatocyte level approaching the maximum liver volume.
This is in line with the notion of administering high dosage early in the treatment \cite{Stengel08} and that the dosage should decrease once the results are
visible. Usually the detection level of virions for HCV patients is taken to be of the order $\sim 10^{2}$ HCV RNA copies per ml \cite{Dixit08}.
We define $t_{d}$ to be the time at which the virion levels become undetectable.
As seen in Figure \ref{three_state}, the viral load drops below this level for $u^{*}$ after about $6$ weeks (\textit{i.e,} $t_{d}=6$ weeks).
We then examine the state dynamics of a large number of sets of sample parameter values under this
optimal control $u^{*}$.
The sample parameter values are generated using the Latin Hypercube Sampling technique.
We implement this sampling method to generate a large number of sample points for all parameter values from
a multivariate distribution (normal, in our case, with a specified mean vector $\stackrel{\rightarrow}{\mu}$ and covariance matrix $\Sigma$)
\cite{Pachpute11}. For the theoretical and computational aspects of this method, the interested reader may
refer to the book by Glasserman \cite{Glasserman04}. We take the mean vector  $\stackrel{\rightarrow}{\mu}$ to be the values
specified in Table \ref{tableone}, \textit{i.e.,} $\mu=\left(s,d,p,\beta,c,\delta,r,T_{\max}\right)$.
The diagonal covariance matrix $\Sigma$ is taken as $\text{diag}\left(s^2,d^2,p^2,\beta^2,c^2,\delta^2,r^2,T_{\max}^{2}\right)$ \cite{Pachpute11}.
Once the sets of parameter values are generated they are tested for positivity and physiological conditions
($r>d$ and $s\le d T_{\max}$ \cite{Dahari07}) till exactly $500$ such sets are accepted.
The sample sets, which failed this test, are rejected. We, then, examine the dynamics of the
state variables (by numerically approximating Equation (\ref{model1eqnnew})) for all the accepted sample parameter sets under the application of the
optimal therapy $u^{*}$. We keep track of the time ($t_{d}$) required for the viral load to fall below detection level of $10^{2}$ for each sample set.
The cumulative distribution of $t_{d}$ is shown in Figure \ref{three_distribution}.
The figure shows that, about half of the cases exhibit a short-term response to the optimal therapy, which obviously includes the acute cases,
where the virus clears out irrespective of whether any treatment is administered or not \cite{Pachpute11}.

We also perform Spearman's test to check for monotonic relations between each of the sample parameters sets and the time required ($t_{d}$)
for the viral load to drop below detection level. This test indicates a strong negative correlation between $t_{d}$ and the death rate of infected hepatocytes $\delta$.
A reason for this could be the strong negative correlation between $\delta$ and the transcritical bifurcation ($C_{1}^{*}$) over the same distribution \cite{Pachpute11}.
Biologically, the cases, which present with a higher death rate of infected hepatocytes, are more likely (as compared to other parameters)
to respond to therapy. Also, $t_{d}$ shows a mild monotonic correlation with $r$. A higher rate of proliferation results in higher infected
hepatocyte levels, which in turn supports the production of virions. Therefore, the cases which present with a high rate of proliferation
might not achieve short-term response.

\section{Model for Combination Treatment with IFN and Ribavirin}

In this section, we present an extended model which includes the therapeutic effects
of ribavirin to the model discussed in Section 2. This model, for combination treatment with IFN and ribavirin,
comprises of four coupled ODEs:
\begin{eqnarray}
\label{model2eqn}
\frac{dT}{dt}&=&s+rT\left(1-\frac{T+I}{T_{\max}}\right)-dT-\beta T V_{I} \nonumber\\
\frac{dI}{dt}&=&\beta TV_{I}+rI\left(1-\frac{T+I}{T_{\max}}\right)-\delta I \nonumber\\
\frac{dV_{I}}{dt}&=&(1-\rho)(1-\epsilon_{p})p I-cV_{I}\nonumber\\
\frac{dV_{NI}}{dt}&=&\rho(1-\epsilon_{p})p I-cV_{NI}.
\end{eqnarray}
Ribavirin works by rendering a fraction of the newly produced virions non-infectious.
The virion population is divided into two different virion populations, namely
infectious ($V_{I}$) and non-infectious ($V_{NI}$), as a result of administration of ribavirin
with an efficacy of $\rho$. This model also admits two steady states \cite{Dahari07}, namely the uninfected state,
\[T^{(u)}= \frac{T_{\max}}{2r}\left[r-d+\sqrt{(r-d)^{2}+\frac{4rs}{T_{\max}}}\right],
I^{(u)}=0, V_{I}^{(u)}=0,V_{NI}^{(u)}=0.\]
and the infected state,
\begin{eqnarray*}
T^{(i)}&=&\frac{1}{2}\left[-D+\sqrt{D^{2}+\frac{4sT_{\max}}{rA^{2}}}\right],I^{(i)}=T^{(i)}(A-1)+T_{\max}-B\\
V_{I}^{(i)}&=&\frac{(1-\epsilon_{p})(1-\rho)pI^{(i)}}{c},V_{NI}^{(i)}=\frac{\rho V_{I}^{(i)}}{(1-\rho)},
\end{eqnarray*}
where,
\[A=\frac{(1-\epsilon_{p})(1-\rho)p\beta T_{\max}}{cr}, B=\frac{\delta T_{\max}}{r},
D=\frac{1}{A}\left[T_{\max}+\frac{dB}{\delta A}-B\left(\frac{1}{A} + 1\right)\right].\]
Under the physiological conditions $r>d$ and $s\le dT_{\max}$, it can be shown \cite{Dahari07}
that there is a transcritical bifurcation at,
\[(1-\epsilon_{p})(1-\rho)=\frac{c(\delta T_{\max}+rT^{(u)}-rT_{\max})}{p\beta T^{(u)}T_{\max}}=C_{2}^{*} (\text{say}).\]
In other words, the uninfected steady state is stable if $(1-\epsilon_{p})(1-\rho)<C_{2}^{*}$ while the
infected steady state is stable if $(1-\epsilon_{p})(1-\rho)>C_{2}^{*}$.

\subsection{Optimal Control Problem for Combination Treatment with IFN and Ribavirin}

We define a slightly modified objective functional for the four equation model outlined in the previous section.
In defining the objective functional, we consider the integral values of the infected hepatocyte concentration ($I$) and infectious viral
load ($V_{I}$), as well as the terminal values of infected hepatocytes and both the virion populations ($V_{I}$ and $V_{NI}$).
The integral value of the non-infectious viral load is not taken into account (\textit{i.e.,} has zero cost coefficient), since
it does not contribute towards the infectivity of hepatocytes during the course of the treatment. Also, the functional incorporates
the need to minimize the therapeutic side-effects of both IFN and ribavirin.
Under these biomedical considerations, the objective functional for combination therapy is defined as,
\begin{eqnarray*}
\Jc&=&\frac{1}{2}\left[S_{22}I(t_{f})^2+S_{33}V_{I}(t_{f})^2+S_{44}V_{NI}(t_{f})^2 \right]\\
&+&\frac{1}{2}\int_{t_{i}}^{t_{f}}\left[Q_{22}I(t)^2+Q_{33}V_{I}(t)^2+R_{11}\epsilon_{p}(t)^{2}+R_{22}\rho(t)^{2}\right]dt,
\end{eqnarray*}
The state vector for this model is defined as
$\xbf=\begin{pmatrix}x_{1}&x_{2}&x_{3}&x_{4} \end{pmatrix}^{\top}$, where $x_{1}=T, x_{2}=I, x_{3}=V_{I}$ and $x_{4}=V_{NI}$.
Similarly, the control vector is defined as $\ubf=\begin{pmatrix} u_{1}&u_{2} \end{pmatrix}^{\top}$,
where $u_{1}=\epsilon_{p}$ and $u_{2}=\rho.$
The system described in Equation (\ref{model2eqn}) is written as $\dot{\xbf}=f(\xbf(t),\ubf(t),t)$
with the initial condition $\xbf(t_{i})=\xbf_{0}$, which, as in case of the first model, is taken to be the pre-treatment ($u_{1}=u_{2}=0$)
infected steady state concentrations. The subsequent mathematical and numerical treatment of the problem is analogous
to the approach adopted for the monotreatment model.

\subsection{Results and Discussions for Combination Treatment with IFN and Ribavirin}

As in the case of monotreatment, the parameter values are taken from Table \ref{tableone}.
The optimal control problem is solved using the steepest gradient method described in Section $2$.
The cost coefficients used for numerical implementation  are $S_{22}=S_{33}=S_{44}=10^{-5}$, $Q_{22}=Q_{33}=10^{-5}$ and $R_{11}=R_{22}=0.1$.

\begin{figure}[hb]
\centering
\includegraphics[width=0.7\textwidth]{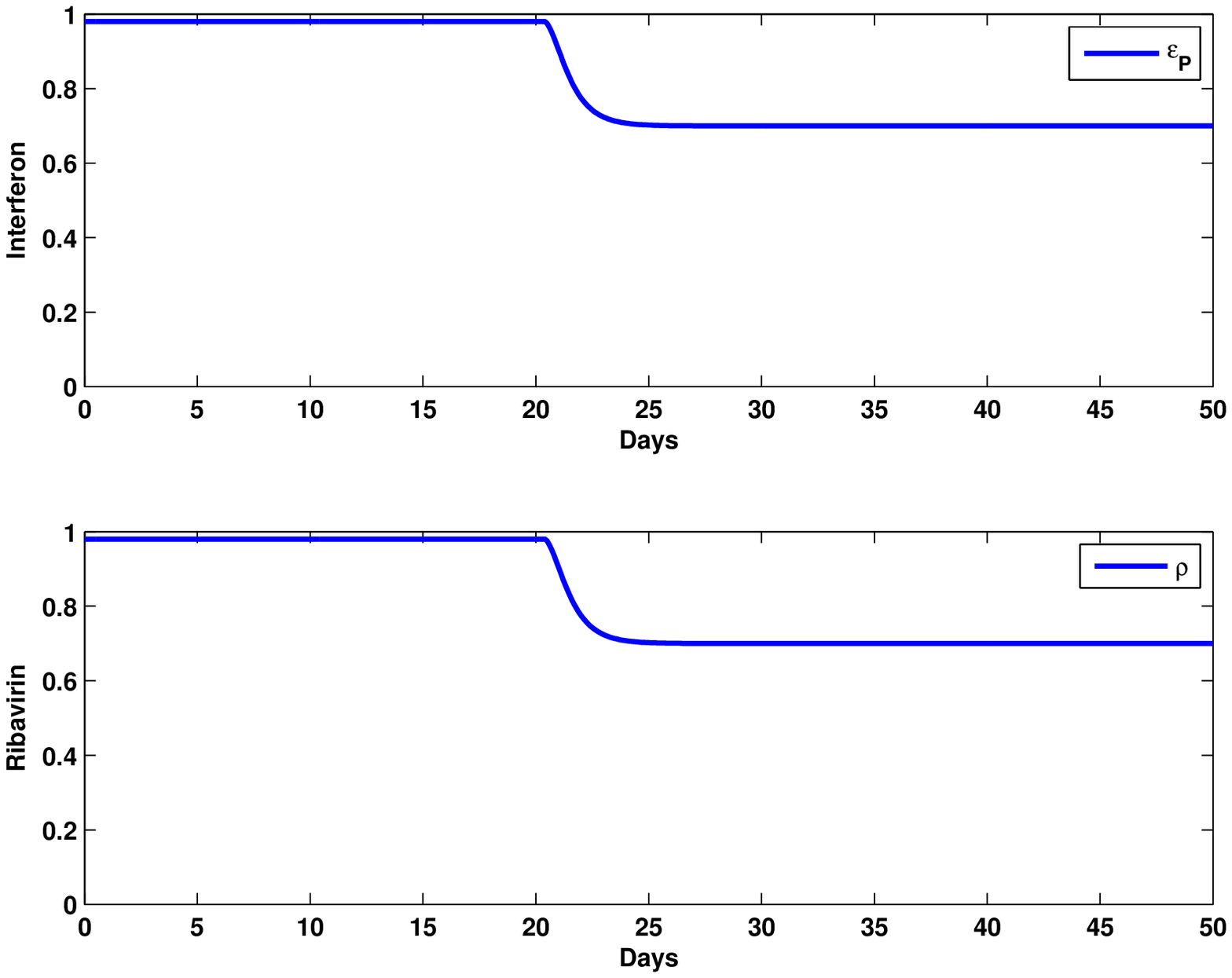}
\caption{Optimal Control for Combination Therapy}
\label{four_control}
\end{figure}

\begin{figure}[hb]
\centering
\includegraphics[width=0.7\textwidth]{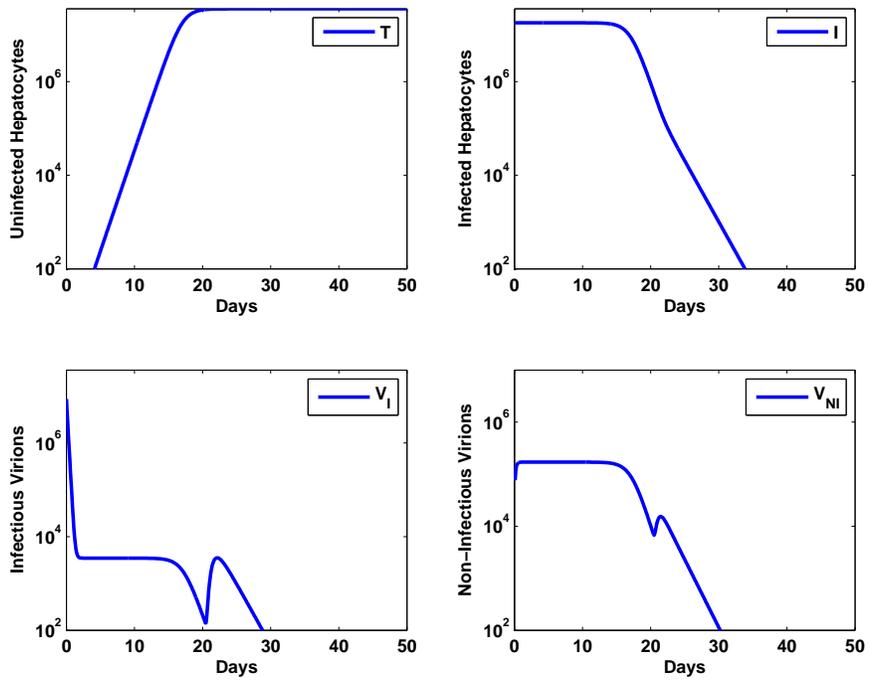}
\caption{State Dynamics for Combination Therapy}
\label{four_state}
\end{figure}

The optimal combination therapy $\ubf^{*}$ and the dynamics of the state variables under $\ubf^{*}$
are presented in Figure (\ref{four_control}) and (\ref{four_state}) respectively.
As it can be seen from the figures, the patient response, as a result of combination therapy, is better
as compared to monotreatment regimen.

It is observed from Figure (\ref{four_state}) that, the viral load drops below the detection level ($10^{2}$) after about $30$ days of combination
treatment as compared to $40$ days in case of monotreatment. In addition, decline in infected hepatocyte levels in this case is steeper,
in comparison with the decline under monotherapy.
The optimal control remains at $\overline{\ubf}=(0.98,0.98)^{\top}$ for about three weeks and then gradually decreases over the rest of the period,
as it was seen in case of monotherapy as well. If the treatment window is varied, length of this period of high efficacy remains
roughly unchanged for smaller values of $R$, which reiterates the point made earlier that, in case of drugs with minor side effects,
the duration of high dosage remains approximately the same. As was the case with monotherapy, the infectious viral load ($V_{I}$)
shows a triphasic decline during the period of administering high dosage, while the uninfected hepatocytes ($T$) exhibit an increase
in levels. Beyond this period, the uninfected hepatocyte count remains close to $T_{\max}$. The infected hepatocyte count ($I$) shows a
concurrent constant level followed by a decline. This observation is consistent with the idea that, high doses are administered until
visible physical progress is achieved, after which, dosage is reduced.

\begin{figure}[hb]
\centering
\includegraphics[width=0.7\textwidth]{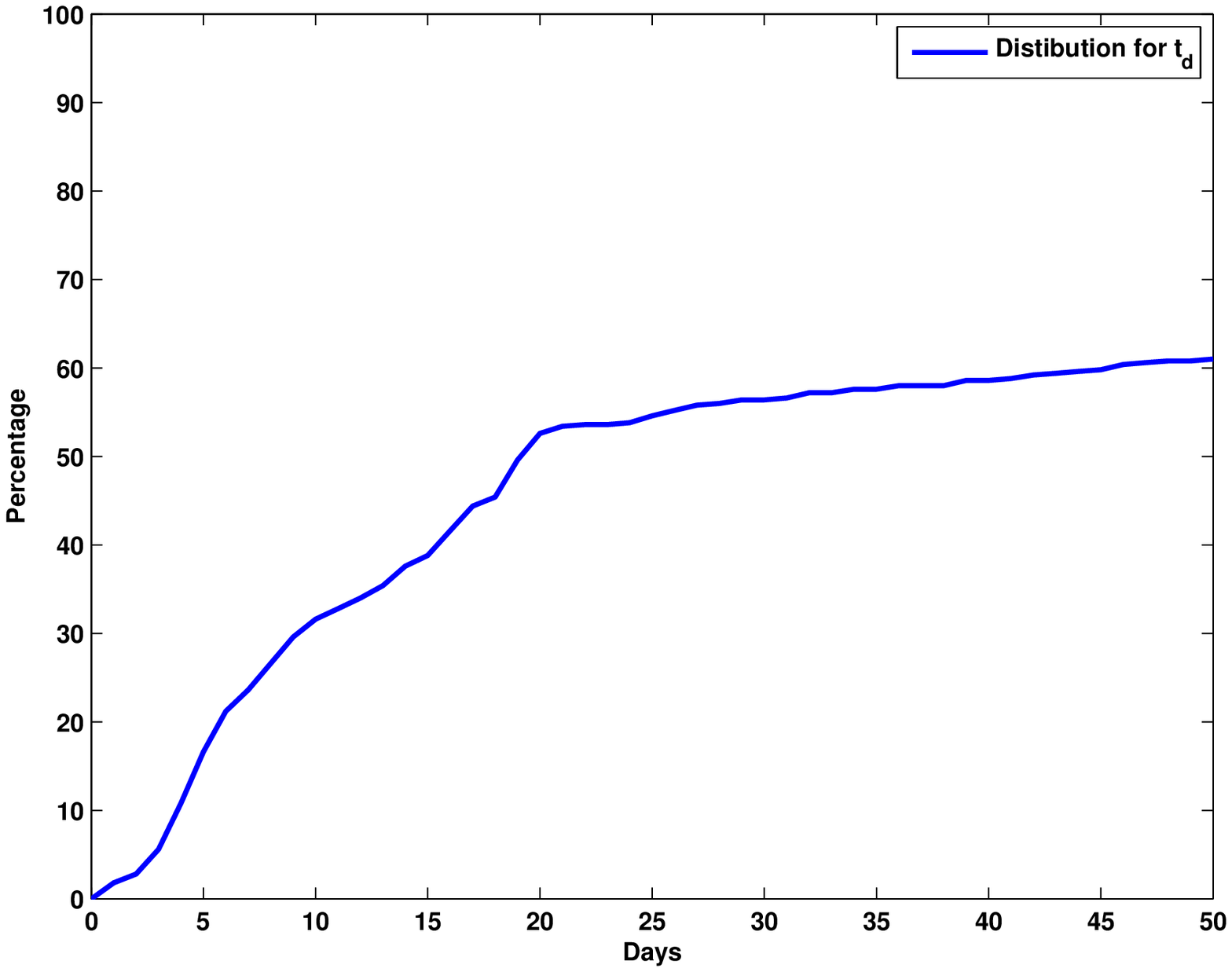}
\caption{Distribution of $t_{d}$ for Combination Therapy}
\label{four_distribution}
\end{figure}

We use the Latin hypercube sampling technique described in Section $3$ to generate sample parameters, using the same
$\stackrel{\rightarrow}{\mu}$ and $\Sigma$ as before along with
positivity and physiological restrictions \cite{Dahari07} for this model as conditions for acceptability.
The cumulative distribution for $t_{d}$ is determined for the accepted sample parameter sets
and is shown in Figure (\ref{four_distribution}). The graph for this distribution shows an improvement in patient response ($60\%$ after
50 days) under combination therapy, as compared to monotreatment success rate of about $50\%$.
Spearman's tests, between sample parameters and $t_{d}$, show that a strong negative correlation exists between $\delta$ and
$t_{d}$. This is consistent with our earlier work \cite{Pachpute11}, where $\delta$ was found to be the most crucial parameter
in determining the patient response. Also, a moderate positive correlation was observed between the rate of proliferation $r$ and $t_{d}$.
A higher proliferation of infected hepatocytes supports the replication of HCV, a result also noted in \cite{Pachpute11}.

\section{Conclusion}

In this paper, we consider two well established models for HCV dynamics, incorporating the efficacy of the drugs administered.
The effects of the drugs as manifested by this model, vary greatly amongst individual patents, depending upon the parameter
variation. A transcritical bifurcation condition which identifies the stability criterion for the uninfected and infected
steady state is presented. Optimal treatment protocol is determined for both the models, one for monotreatment with IFN and the other for a
combination therapy of IFN and ribavirin. The optimal therapy shows an initial high level followed by a decline  once the hepatocyte gradually gets restored.
The results, obtained in this paper, are consistent with biomedical observations relating therapeutic protocols, especially, the notion of administering
high dosage in the early stages of infection and gradual lowering of the dosage after the patient has shown signs of recovery.

With the goal of examining the consequences of administering the optimal efficacy of the two drugs in case of several
patient scenarios, we generate a large set of patient parameters using Latin hypercube sampling.
The state dynamics for the generated parameter sets are studied under the optimal therapy. The results indicated a better response
in case of combination therapy as compared to monotherapy. Statistical tests, presented in this paper, show that, patient
recovery is highly influenced by the death rate of infected hepatocytes. Also a higher proliferation rate abets the viral replication.

Such statistical analysis can be helpful in determining the optimal therapy on an individual basis,
by recognizing the major factors which influence patient response.
We believe that the analysis presented in this paper, combined with pharmacokinetics studies, could play an important role
in developing improved HCV treatment regimen.


\begin{thebibliography}{99}

\bibitem {Dixit08} Dixit NM (2008) Advances in the mathematical modelling of hepatitis C virus dynamics.
Journal of the Indian Institute of Science 88(1):37–43.

\bibitem{Roe08} Roe B, Hall WW (2008) Cellular and molecular interactions in coinfection with hepatitis C virus and human immunodeficiency virus.
Expert Reviews in Molecular Medicine 10: null-null.

\bibitem{Razali07}
Razali K. et al. (2007) Modelling the hepatitis C virus epidemic in Australia.
Drug and Alcohol Dependence 91: 228-235.

\bibitem{Castello10}
Castello G, Scala S, Palmieri G, Curley SA, Izzo F (2010), HCV-related hepatocellular carcinoma: From chronic inflammation to cancer.
Clinical Immunology 134: 237-250.

\bibitem{Pal02} Pal SK, Chalamalasetty BK, Choudhuri G (2002) Hepatitis C: a major health problem of India.
Current Science 83(9): 1058-1059.

\bibitem{Mukhopadhya08} Mukhopadhya A (2008) Hepatitis C in India.
Journal of Biosciences 33(4): 465-473.

\bibitem{Neumann98} Neumann AU, Lam NP, Dahari H, Gretch DR, Wiley TE, Layden TJ, Perelson AS. (1998)
Hepatitis C viral dynamics in vivo and the antiviral efficacy of interferon-$\alpha$ therapy. Science 282: 103-107.

\bibitem{Dixit04} Dixit NM, Layden-Almer JE, Layden TJ, Perelson AS (2004) Modelling how ribavirin improves interferon response rates
in hepatitis C virus infection. Nature 432:922–924.

\bibitem{Dahari07} Dahari H, Lo A, Ribeiro RM, Perelson AS (2007) Modeling hepatitis C virus dynamics: liver regeneration and critical drug efficacy.
Journal of Theoretical Biology 247: 371-381.

\bibitem{LenhartBook} Lenhart S, Workman JT (2007), Optimal control applied to biological methods.

\bibitem{DePillis07} De Pillis LG, Gu W, Fister KR, Head T, Maples K, Murugan A,
Neal T, Yoshida K (2007), Chemotherapy for tumors: An analysis
of the dynamics and a study of quadratic and linear optimal controls,
Mathematical Biosciences, 209(1), 292-315.

\bibitem{Fister03} Fister KR, Panetta JC (2003), Optimal control applied to competing
chemotherapeutic cell-kill strategies , SIAM Journal on Applied Mathematics, 63(6): 1954-1971

\bibitem{Murray90} Murray JM (1990), Some optimal control problems in cancer chemotherapy with a toxicity limit,
Mathematical Biosciences, 100(1): 49-67

\bibitem{Chavez09} Chavez IYS, Morales-Menendez R, Chapa SOM (2009), Glucose optimal control system in diabetes treatment ,
Applied Mathematics and Computation, 209(1): 19-30

\bibitem{Kirschner97} Kirschner D, Lenhart S, Serbin S (1997), Optimal control of the chemotherapy of HIV,
Journal of Mathematical Biology, 35(7): 775-792

\bibitem{Joshi02} Joshi HR (2002), Optimal control of an HIV immunology model,
Optimal Control Applications and Methods, 23(4): 199-213

\bibitem{Adams05} Adams BM, Banks HT, Davidian M, Kwon H-D, Tran HT,
Wynne SN, Rosenberg ES (2005) HIV dynamics: Modeling, data analysis, and optimal
treatment protocols, Journal of Computational and Applied Mathematics 184:10-49

\bibitem{Stengel08} Stengel RF (2008),
Mutation and control of the human immunodeficiency virus, Mathematical Biosciences, 213(2): 93-102

\bibitem{SPCJBS} Chakrabarty SP, Joshi HR (2009),
Optimally controlled treatment strategy using interferon and ribavirin for hepatitis C,
Journal of Biological Systems, 17(1): 97-110

\bibitem{SPCOCAM} Chakrabarty SP (2009), Optimal efficacy of ribavirin in the treatment of hepatitis C,
Optimal Control Applications and Methods, 30(6): 594-600

\bibitem{Martin11}
Martin NK, Ashley B, Pitcher AB, Vickerman P, Vassal A, Hickman M (2011),
Optimal control of hepatitis C antiviral treatment programme delivery for prevention amongst a population of injecting drug users,
PLos One, 6(8): e22309.

\bibitem{Kirk04} Kirk DE (2004), Optimal control theory : An Introduction, Dover Publications.

\bibitem{Powers03} Powers KA, Dixit NM, Ribeiro RM, Golia P, Talal AH, Perelson AS (2003),
Modeling Viral and Drug Kinetics: Hepatitis C Virus Treatment with Pegylated Interferon Alfa-2b,
Seminars in Liver Disease, 23: 13-18.

\bibitem{Pachpute11} Pachpute G, Chakrabarty SP (2011), Analysis of hepatitis C viral dynamics using Latin Hypercube Sampling,
\url{http://arxiv.org/abs/1106.5313}

\bibitem{Glasserman04} Glasserman P (2004), Monte Carlo methods in financial engineering, Springer.

\end{thebibliography}
\end{document}